# The alignment of nematic liquid crystal by the Ti layer processed by nonlinear laser lithography

Ihor Pavlov

*Department of Physics, Bilkent University, Ankara, Turkey*

Bilkent University, Ankara, 06800, Turkey,

*Department of Photon Processes, Institute of Physics of the National Academy of Sciences of Ukraine, Kyiv, Ukraine*

Institute of Physics of the National Academy of Sciences of Ukraine, Prospekt Nauki 46, Kyiv-28, 03028, Ukraine, E-mail: pavlov.iop@gmail.com

Andrey Rybak

*Department of Photon Processes, Institute of Physics of the National Academy of Sciences of Ukraine, Kyiv, Ukraine*

Institute of Physics of the National Academy of Sciences of Ukraine, Prospekt Nauki 46, Kyiv-28, 03028, Ukraine, E-mail: rybak.andrey@gmail.com

Andrii Dobrovolskiy

*Department of Gas Electronics, Institute of Physics of the National Academy of Sciences of Ukraine, Kyiv, Ukraine*

Institute of Physics of the National Academy of Sciences of Ukraine, Prospekt Nauki 46, Kyiv-28, 03028, Ukraine, E-mail: dobr@iop.kiev.ua

Viktor Kadan

*Department of Photon Processes, Institute of Physics of the National Academy of Sciences of Ukraine, Kyiv, Ukraine*

Institute of Physics of the National Academy of Sciences of Ukraine, Prospekt Nauki 46, Kyiv-28, 03028, Ukraine, E-mail: kadan@iop.kiev.ua

Ivan Blonskiy

*Department of Photon Processes, Institute of Physics of the National Academy of Sciences of Ukraine, Kyiv, Ukraine*

Institute of Physics of the National Academy of Sciences of Ukraine, Prospekt Nauki 46, Kyiv-28, 03028, Ukraine, E-mail: blon@iop.kiev.ua

Fatih Ö. Ilday

*Department of Physics, Bilkent University, Ankara, Turkey*

*Department of Electrical and Electronics Engineering, Bilkent University, Ankara, Turkey*

*UNAM - National Nanotechnology Research Center and Institute of Materials Science and Nanotechnology, Bilkent University, Ankara, Turkey*

Bilkent University, Ankara, 06800, Turkey, E-mail: ilday@bilkent.edu.tr

Zoya Kazantseva

*Electrical and Galvanomagnetic Properties of Semiconductors, V.E. Lashkaryov Institute of Semiconductor Physics of the National Academy of Sciences of Ukraine, Kyiv, Ukraine*

V.E. Lashkaryov Institute of Semiconductor Physics of the National Academy of Sciences of Ukraine, Prospekt Nauki 41, Kyiv-28, 03028, Ukraine, E-mail: kazants@isp.kiev.ua

Igor Gvozdovskyy[*] (corresponding author)

*Department of Optical Quantum Electronics, Institute of Physics of the National Academy of Sciences of Ukraine, Kyiv, Ukraine*

Institute of Physics of the National Academy of Sciences of Ukraine, Prospekt Nauki 46, Kyiv-28, 03028, Ukraine, telephone number: +380 44 5250862, [*]E-mail: igvozd@gmail.com

# The alignment of nematic liquid crystal by the Ti layer processed by nonlinear laser lithography

It is well known that the alignment of liquid crystals can be realized by rubbing or photoalignment technologies. Recently nonlinear laser lithography was introduced as a fast, relatively low-cost method for large area nano-grating fabrication based on laser-induced periodic surface structuring. In this letter for the first time the usage of the nonlinear laser lithography as a perspective method of the alignment of nematics was presented. By nonlinear laser lithography, microgrooves with about 1 μm period were formed on Ti layer. The microstructured Ti layer was coated with oxidianiline-polyimide film with annealing of the polymer followed without any further processing. Aligning properties of microstructured Ti layers were examined with combined twist LC cell. The dependencies of the twist angle of LC cells and azimuthal anchoring energy of layers on scanning speed and power of laser beam during processing of the Ti layer were the focus of our studies as well. The maximum azimuthal anchoring energy, obtained for pure microstructured Ti layer, is comparable with photoalignment technology. It was found that the deposition of polyimide film on microstructured Ti layer leads to the gain effect of the azimuthal anchoring energy. Also, AFM study of aligning surfaces was carried out.

**Introduction**

The alignment of liquid crystals (LCs) is important and a key condition for their application as a liquid with anisotropic properties. For this purpose both different aligning materials and various methods of their processing can be used, as demonstrated, for instance, in the book [1] and many reviews. [2-5] However, in present two main methods can be emphasized, which are used for the creation of the aligning layers and were well studied for further application in display technology, *etc*. The first method, used extensively for different applications in industry, is the rubbing technique with various materials [1,6-9] of different surfaces. [1,10] Despite the fact that the rubbing technology is widely used in LCD technology, however this technique has some shortcomings, among which are accumulation of both the static charges and dust particles. [4] The other method of the homogeneous aligning of LCs is a recently discovered photoalignment effect. [11-15] It is a real alternative method to the rubbing technique, because the usage of photosensitive materials, deposited on a substrate or dissolved in the bulk of LCs, leads to the change of the orientational order of photoproducts under polarized light irradiation. In this case the mechanical contact with the surface of a substrate, with all the shortcomings that this entails,[4] is absent. However, photoaligning materials often possess photodegradation, and in order to create a relatively large area of aligning surfaces there is a need for a long-term exposure and preparation of masks. As it was noted in Ref. [4], the photoalignment technology gives an effective control of main anchoring parameters, namely: easy orientation axis, pretilt angle and anchoring energy (AE). It is noted that depending on chemical properties of materials, [2,4] AE values of the photoaligning surfaces can be within a wide range $10^{-8} - 10^{-5}$ J/m$^2$.

In the case of both methods, the formation of the aligning layers can be created by means of deposition of inorganic materials, by polymer-coating from different solutions, by using of Langmuir-Blodgett, spin-coating and dipping techniques, followed by polymerization and their further processing with application of a rubbing roll or polarized light. In addition, it should be noted that the usage of plasma beam as a method of processing aligning layers for the homogeneous orientation of LC was recently studied. [16,17] After a certain method of the processing of the aligning layers, it was observed that the period of the grooves structure can change within a range 100 – 300 nm, while the amplitude (depth) of relief can be within a range 80 - 150 nm. [6,7,18,19] The homogeneous alignment of the LC cell was observed due to, on the one hand, the appearance of the anisotropic properties of aligning layers and, on the other hand, long-range interaction of the LC molecules in a mesophase.

In addition, to obtain the LC alignment by a surface with a small period (about 235 -250 nm) of nanogrooves the e-beam lithography [20] and AFM nano-rubbing [21] were applied. However, the area of nanogrooves rubbed on the surface was very small, [20] and both methods show a very low throughput. Moreover, the nano-imprint lithography [22] and photolithography [23] were also used to create nanogrooves on a polymer surface for aligning LCs, but both techniques are complicated due to the preparation of masks.

It should be also noted that recently, the fast and high-throughput method, consisting of the splitting of a polymer film with further propagation of the wave front to induce self-assembled micro- and nanogrooves on a polymer surface (the so-called crack-induced grooving method), was proposed to align LCs.[24]

The possibility of processing of the various materials by ultrafast laser pulses for different applications is described in detail. [25] In this letter, for the first time we have made aligning layers, using a simple, relatively low-cost and high-speed method of the creation of a large area of high anisotropic metal-oxide periodic microstructures (the so-called ripples) with femtosecond laser. [26] Conditionally, it can be done in one or two stages. Contrary to the traditional methods [1-19] of the LC alignment, at the first stage the processing of titanium (Ti) layer, deposited on a glass substrate, which results in the periodic microgrooves with certain parameters (for instance, depth, period and angle of direction of ripples), is realized. At the second stage, the creation of aligning surfaces owing to the coating of a polymer onto a layer with ~ 1 μm period of microgrooves without additional processing of a polymer (such as rubbing or irradiation with polarized light), excepts the process of polymerisation,. The calculation of AE of both the pure microstructured Ti layer (MSTL) and similar layer with the polymer-coated surface were carried out on the basis of the measurement of the twist angle in the combined twist LC cell. [27,28] The dependence of the calculated azimuthal AE on the scanning speed and laser pulse fluence (LPF) during periodic Ti surface structuring was studied.

**Experiment**

To measure the twist angle and calculate AE herein the idea of combined twist LC cells, consisting of the tested and reference substrates was used. [27-29]

The tested substrates are used of two types. The first type of the tested substrate was coated with 300 nm Ti layer and further processed by the method of nonlinear laser lithography (NLL), using the laser scanning with femtosecond pulse [26] ($P$ = 350 mW, $f$ = 1 MHz and $I$ = 350 nJ). The direction of microstructures was inclined at a fixed angle of 8º to the one side of the substrate.

The second type of the tested substrate consists of the first type substrate additionally coated with 1-% DMF solution of oxidianiline-polyimide (ODAPI) (Kapton synthesized by I. Gerus, Ukraine) by means of dipping technique using equipment for Langmuir-Blodgett film preparation R&K (Wiesbaden, Germany). The second type of tested substrate was dipping into appropriable solution and further vertical drawing up at constant speed about 5 mm/min along direction of the microstructured Ti layers.

The reference substrate was coated with polyimide PI2555 (HD MicroSystems, USA) processed with the rubbing technique having strong AE W = (4 ± 1)×10$^{-4}$ J/m$^2$ for the certain pressure and number of times of unidirectional rubbings. [30,31] The rubbing was carried out at 45 degrees to one side of the substrate. The thickness of a gap was set to 20 – 25 μm by a Mylar spacer and measured by the interference method, using transmission spectra of empty LC cells. The LC cells were filled with the nematic LC E7 (obtained by Licrystal Merck, Germany) with twist elastic constant $K_{22}$ = 6.8 pN, [32] at the higher temperature T = 61 $^o$C than the temperature of the isotropic phase ($T_{Iso}$ = 58 $^o$C) and slowly (~ 0.1$^o$ C/min) cooled.

To measure the twist angle $\varphi$ the simple scheme described in [29] was used. For this the output monochromatic linear-polarized light ( λ = 632.8 nm and power about 1.5 mW) from He-Ne laser LGN-207a (Lviv, Ukraine) and a silicon photodetector (PD) FD-18K (Kyiv, Ukraine) with the spectral range 470 ÷ 1100 nm were chose. PD was connected to the oscilloscope Hewlett Packard 54602B 150MHz (USA).

**Results and discussion**

Figure 1 (a) presents the SEM image of the pure MSTL with a high degree of anisotropy. The dependence of the microgrooves depth A on LPF $\mathscr{I}$, showed in Figure 1 (b), has analogy with the results in Ref. [19], where the influence of exposure on the value of the grooves depth during the photoaligning technique was studied. The cross section of microgrooves of the pure MSTL obtained by AFM is shown in Figure 1 (c). At constant parameters of the NLL ($\mathscr{I}$ = 0.55 J/cm$^2$ and $\upsilon$ = 1500 mm/s) by AFM study it is found that the certain periodic structure with an average period Λ = 0.92 μm and average depth A = 225 nm is formed. In contrast to methods traditionally used for the alignment of LCs, [6,7,18,19] where the period of microgrooves is within a range 100 - 300 nm, it can be assumed that owing to anisotropy of MSTLs the homogeneous alignment of the nematic LC may be observed, however, AE of the layers will be lower.

To estimate the order of the AE value of pure MSTLs we use the Berreman's theory. [6] According to this theory, the AE depends on the depth A and period Λ of microgrooves, and can be written as: [6,18,19]

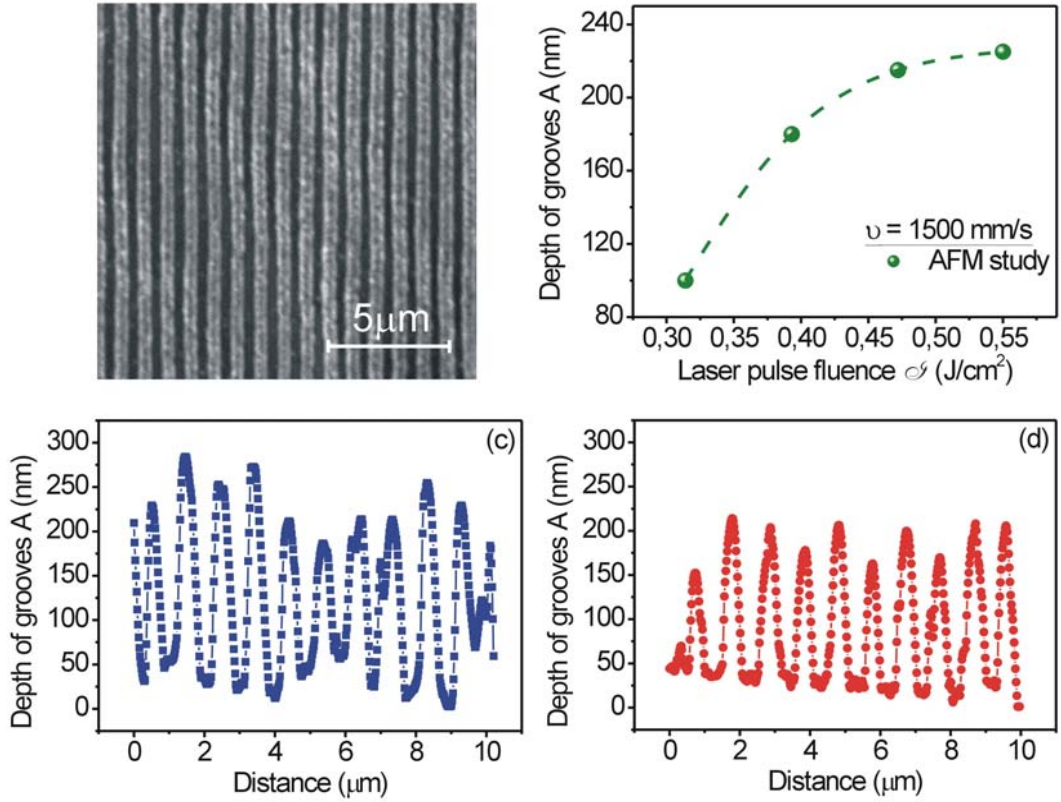

Figure 1. (a) SEM image of the pure MSTL after processing with the NLL method. (b) Dependence of the average depth A on the LPF $\mathscr{I}$ of the pure MSTL after the NLL processing. The cross section of microgrooves of the Ti layer: (c) pure, possessing the average period $\Lambda = 0.92$ μm and average depth $A = 225$ nm and (d) coated with the ODAPI film, having the average period $\Lambda = 0.9$ μm and average depth $A = 175$ nm. In the NLL method the LPF $\mathscr{I}$ and scanning speed $\upsilon$ was 0.582 J/cm$^2$ and 1500 mm/s, respectively.

$$W_B = 2\pi^3 \times A^2 \times K_{22} / \Lambda^3 \qquad (1)$$

Thus, by using both period and depth of microgrooves measured by AFM, the value $W_B$ of the pure MSTL reaches $\sim 2.7 \times 10^{-5}$ J/m$^2$. This value is of order of the azimuthal AE of azopolymers, [2,33] photo-crosslinking materials, [2,5,34-36] and polyvinylcinnamate (PVCN). [35,36] According to the theory, [6] to increase the AE value of the pure Ti layer, it is necessary either to decrease the period, or to increase the depth of the microgrooves.

However, here to increase the AE value we propose not to change the main parameter, such as the wavelength of a laser, resulting in the change of the microgrooves period, during the surface structuring with the NLL method. But we suggest additionally apply a coating of the ODAPI film onto the surface of the MSTL. As can be seen from Figure 1 (d), the average period and average depth of microgrooves was about 0.92 μm and 175 nm, respectively. According to Berreman's theory,[6] the calculated AE value decreased to $1.7 \times 10^{-5}$ J/m$^2$, owing to the insignificant change of the average depth of microgrooves after the coating of the polymer film (Figure 1 (d)).

Let us consider the real AE value measured experimentally, using the method of the combined twist LC cell. [27-29]

The preparation of the combined twist LC cells, consisting of the tested substrate, having MSTLs of both types, and reference substrate with a rubbed PI2555 layer, and measurement of the twist angle $\varphi$ of the sample allow us to calculate the AE value ($W_\varphi$) of the aligning layer of the tested substrates. According to Ref. [27,28] the twist angle $\varphi$ is related to the azimuthal AE $W_\varphi$ as follows:

$$W_\varphi = K_{22} \times \frac{2 \times \sin(\varphi)}{d \times \sin 2(\varphi_0 - \varphi)}, \quad (1)$$

where $d$ is the thickness of the LC cell, $\varphi_0 = 36°$ is the angle between the easy axes of the reference and tested substrates and $\varphi$ is the measured twist angle.

Owing to the NLL method, [26] allowing to do changes in a wide range of the scanning speed $v = 500 - 3000$ mm/s and LPF $\mathscr{I}$ within a range $0.236 - 0.59$ J/cm$^2$, dependencies of $\varphi(v)$ and $\varphi(\mathscr{I})$ were studied. For both types of the tested substrates, dependencies $\varphi(v)$ and $\varphi(\mathscr{I})$ are shown in Figure 2(a) and Figure 2(b), respectively. As can be seen from Figure 2, the twist angles of the LC cell are reaching maximum values at a certain range of values of $v = 1300 - 2000$ mm/s or $\mathscr{I} = 0.393 - 0.582$ J/cm$^2$ for both types of the tested substrates.

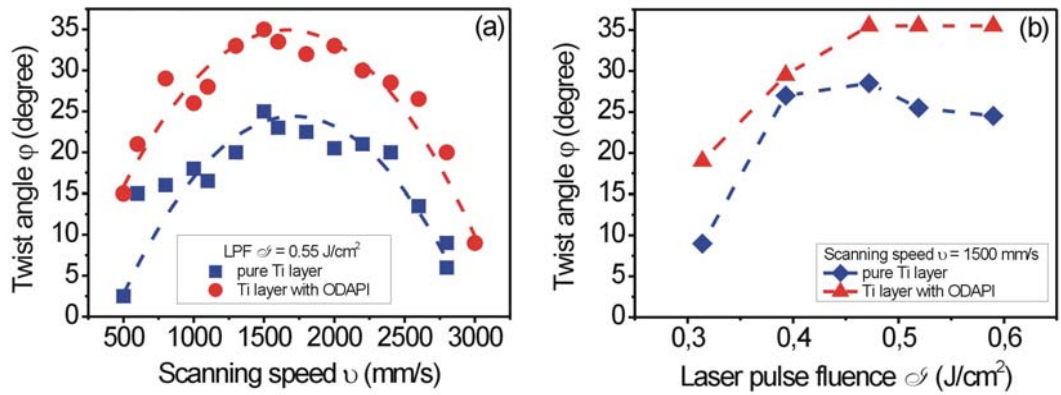

Figure 2. Dependence of the twist angle $\varphi$ of the LC cell on: (a) the scanning speed $v$ at a constant LPF $\mathscr{I} = 0.55$ J/cm$^2$ and (b) the LPF $\mathscr{I}$ at a constant scanning speed $v = 1500$ mm/s during the processing of the Ti layer with the NLL method. The LC cell consists of the tested substrate having the pure MSTL (closed "blue" squares and diamonds) and MSTL coated with the ODAPI film (closed "red" circles and triangles). The dashed line is a guide to the eye.

The dependence of the calculated azimuthal AE of the aligning film on the scanning speed $v$ during the processing of the Ti layer, by using measured twist angles, is shown in Figure 3 (closed "blue" triangles). It is seen that for the pure MSTL the AE reaches the value $\sim 4.6 \times 10^{-6}$ J/m$^2$, which is lower in comparison with the value, obtained by Berreman's theory, [6] but it agrees nicely with the AE of photoaligning layers as was mentioned above. It is evident that the difference between values of anchoring energies, experimentally obtained with the use of the twist LC cell (Equation (2)) and from Berreman's calculation, is observed due to the fact that the theory does not take into account the interaction between the LC molecules and aligning surface. The same conclusion was made for the photoaligning surfaces [37,38].

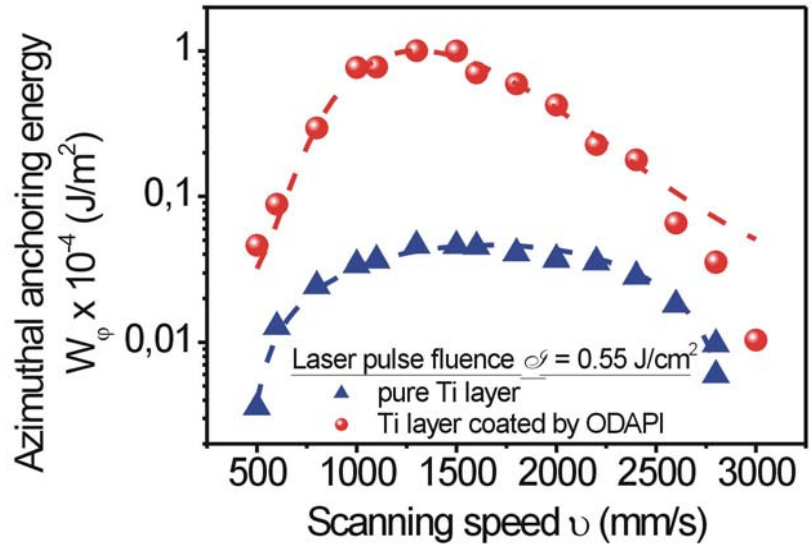

Figure 3. Dependence of the calculated azimuthal AE (Equation 2) of aligning films on the scanning speed $v$ during the processing of the Ti layer with the NLL method. The LC cell consists of the tested substrate having the pure MSTL (closed "blue" triangles) and MSTL coated with the ODAPI film (closed "red" spheres). The dashed line is a guide to the eye.

In the case of the MSTL with the coated ODAPI film the dependence of AE on the scanning speed $v$ is shown in Figure 3 with closed "red" spheres. It is seen that coating of the polymer film onto the MSTL leads to the dramatic increase in the AE value $\sim 1\times10^{-4}$ J/m$^2$. It is obvious that this azimuthal AE consists of at least two components, namely: on the one hand, AE, caused by the microsrtructured layer with a certain period and depth and, on the other hand, owing to physical and chemical properties of the surface of a polymer and its interaction with LC molecules.

In addition, the influence of the LPF $\mathscr{I}$ during the processing of the Ti layer at a certain constant scanning speed $v$ on AE values is shown in Figure 4. As can be seen from Figure 4, for each constant scanning speed $v$ there is an optimal value of the LPF $\mathscr{I}$ when the maximal value of AE is reached. As can be seen form Figure 4(b), the gain of AE to the value $\sim 1\times10^{-4}$ J/m$^2$ can be reached after the processing of the Ti layer with the scanning speed $v$ = 1500 mm/s and LPF $\mathscr{I}$ changing within a range 0.472 – 0.55 J/cm$^2$ and further deposition of the polymer ODAPI. It follows from Figure 4 that during the structuring of the Ti layer the choice of both the scanning speed and LPF allows changing the value of AE within a wide range ~ $10^{-6} – 10^{-4}$ J/m$^2$.

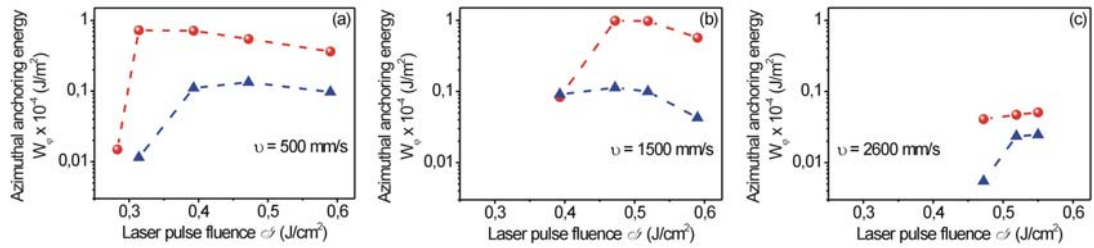

Figure 4. Dependence of the calculated azimuthal AE (Equation 2) of aligning films on the LPF $\mathscr{I}$ at constant scanning speed $v$: (a) 500 mm/s, (b) 1500 mm/s and (c) 2600 mm/s during the processing of the Ti layer with the NLL method. The LC cell consists of the tested substrate having the pure MSTL (closed "blue" triangles) and MSTL coated with the ODAPI film (closed "red" spheres). The dashed line is a guide to the eye.

**Conclusions**

In conclusion to this letter it should be noted that for the first time the NLL method as a new technique of the alignment of nematic liquid crystals was proposed. We are demonstrating the possibility of the AE increasing of the pure Ti layers, owing to the coating of a polymer film onto the microstructured surfaces. It was shown that the pure MSTL is characterized with the relatively weak AE, while the Ti surface with the deposited polymer films has the strong AE. It was shown that the value of AE in a wide range can be controlled by means of the changing at least two parameters (scanning speed and LPF) during the structuring of the Ti layer and further coating of a polymer film without additional processing.

Acknowledgements

The authors thank W. Becker (Merck, Darmstadt, Germany) for his generous gift of nematic liquid crystal E7, Prof. I. Gerus (Institute of Bio-organic Chemistry and Petrochemistry, NAS of Ukraine) for the kind provision of polymer ODAPI and Dr. P. Lytvyn (V. E. Lashkaryov Institute of Semiconductor Physics, NAS of Ukraine) for help during AFM studies. We thank M. Gure and E. Karaman (Bilkent University, Turkey) for technical support of the NLL method. I. P. and F. Ö. I. acknowledge financial support from European Research Council (ERC) Consolidator, Grant "Nonlinear Laser Lithography", No: ERC–617521 NLL.

References:

[1]. Congard J. Alignment of Nematic Liquid Crystals and Their Mixtures London, New York, Paris. Gordon and Breach Science Publishers, 1982.

[2]. Ishimura K. Photoalignment of liquid-crystals systems. Chem Rev. 2000; 100: 1847-1873. Doi: 10.1021/cr980079e

[3]. O'Neill M, Kelly SM. Photoinduced surface alignment for liquid crystal displays. J Phys D: Appl Phys. 2000; 33: R67-R84. Doi: 10.1088/0022-3727/33/10/201

[4]. Yaroshchuk O, Reznikov Yu. Photoalignment of liquid crystals: basic and current trends. J Mater Chem. 2012; 22: 286-300. Doi: 10.1039/C1JM13485J

[5]. Seki T. New strategies and implications for the photoalignment of liquid crystalline polymers. Polymer Journal. 2014; 46: 751-768. Doi: 10.1038/pj.2014.68

[6]. Berreman DW. Solid surface shape and the alignment of an adjacent nematic liquid crystal. Phys Rev Lett. 1972; 28: 1683-1686. Doi: 10.1103/PhysRevLett.28.1683;

[7]. Berreman DW. Alignment of liquid crystals by grooved surfaces. Mol Cryst Liq Cryst. 1973; 23: 215-231.

[8]. Chatelain P. Orientation of liquid crystals by rubbed surfaces. C R Acad Sci. 1941; 231: 875-876.

[9]. Sluckin TJ, Dunmur DA, Stegemeyer H. Crystal That Flow. Classic Papers from History of Liquid Crystals. (Taylor & Francis, London, 2004) p. 723

[10]. Yasutake A, Oshiro RK, France patent application 7536984 (3 December 1975).

[11]. Ishimura K, Suzuki Y, Seki T, et al. Reversible change in alignment mode of nematic liquid crystals regulated photochemically by command surfaces modified with an azobenzene monolayer. Langmuir. 1988; 4: 1214-1216. Doi: 10.1021/la00083a030

[12]. Gibbons WM, Shannon PJ, Sun ST, et al. Surface-mediated alignment of nematic liquid crystals with polarized laser light. Nature. 1991; 351; 49-50. Doi: 10.1038/351049a0

[13]. Schadt M, Shmitt K, Kozinkov V, et al. Surface-induced parallel alignment of liquid crystals by linearly polymerized photopolymers Jpn J Appl Phys. 1992; 31: 2155-2164. Doi: 10.1143/JJAP.31.2155.

[14]. Dyadyusha A, Kozinkov V, Marusii T, et al. Light-induced planar alignment of nematic liquid-crystal by the anisotropic surface without mechanical texture. Ukr Fiz Zh. 1991; 36: 1059-1062.

[15]. Dyadyusha AG, Marusii TYa, Reshetnyak VYu, et al. Orientational effect due to a change in the anisotropy of the interaction between a liquid crystal and a bounding surface. JETF Lett. 1992; 56: 17-21.

[16]. Yaroshchuk O, Kravchuk R, Dobrovolskyy A, et al. Planar and tilted uniform alignment of liquid crystals by plasma-treated substrates. Liq Cryst. 2004; 31: 859-869. Doi: 10.1080/02678290410001703145

[17]. Yaroshchuk O, Zakrevskyy Yu, Dobrovolskyy A, et al. Liquid crystal alignment on the polymer substrates irradiated by plasma beam. In: Klimusheva GV, Iljin AG, editors. Nonlinear Optics of Liquid and Photorefractive Crystals 2000: Proceedings of Eighth International Conference on Nonlinear Optics of Liquid and Photorefractive Crystals. Proceedings of SPIE 4418; 2000 October 2-6; Alushta; Crimea (Ukraine); 2001. Doi: 10.1117/12.428331

[18]. Lee C-R, Fu T-L, Cheng K-T, et al. Surface-assisted photoalignment in due-doped liquid-crystal films. Phys Rev E. 2004; 69: 031704-1-6. Doi: 10.1103/PhysRevE.69.031704

[19]. Fuh AY-G., Liu C-K, Cheng K-T, et al. Variable liquid crystal pretilt angles generated by photoalignment in homeotropically aligned azo dye-doped liquid crystals. Appl Phys Lett. 2009; 95: 161104-1-3. Doi: 10.1063/1.3253413

[20]. Tennant DM, Koch TL, Mulgew PP, et al. Characterization of near-field holography grating masks for optoelectronics fabricated by electron beam lithography. J Vac Sci Technol. B. 1992; 10: 2530-2535.

[21]. Kim JH, Yoneya M, Yakoyama H. Tristable nematic liquid-crystal device using micropatterned surface alignment. Nature. 2002; 420: 159-162. Doi:10.1038/nature01163

[22]. Chou SY, Krauss PR, Renstrom PJ. Imprint lithography with 25-nanometer resolution. Science. 1996; 272: 85-87. Doi: 10.1126/science.272.5258.85

[23]. Hill KO, Malo B, Bilodeau F, et al. Bragg gratings fabricated in monomode photosensitive optical fiber by UV exposure through a phase mask. Appl. Phys. Lett. 1993; 62: 1035-1037. Doi: 10.1063/1.108786

[24]. Lin T-C, Huang L-C, Chou T-R, et al. Alignment control of liquid crystal molecules using crack induced self-assembled grooves. Soft Matt. 2009; 5: 3672-3676. DOI: 10.1039/B911567F

[25]. Malinauskas M, Žukauskas A, Hasegawa S, et al. Ultrafast laser processing of materials: from science to industry. Light Science & Applications. 2016; 5: e16133-13. Doi:10.1038/lsa.2016.133

[26]. Öktem B, Pavlov I, Ilday S, et al. Nonlinear laser lithography for indefinitely large-area nanostructuring with femtosecond pulse. Nat Photonics. 2013; 7: 897-901. Doi: 10.1038/NPHOTON.2013.272

[27]. Andrienko D, Kurioz Yu, Nishikawa M, et al. Control of the anchoring energy of rubbed polyimide layers by irradiation with depolarized UV light. Jpn J Appl Phys. 2000; 39: 1217-1220. Doi: 10.1143/JJAP.39.1217

[28]. Gerus I, Glushchenko A, Kwon S-B, et al. Anchoring of a liquid crystal on photoaligning layer with varying surface morphology. Liq Cryst. 2001; 28: 1709-1713. Doi: 10.1080/02678290110076371

[29]. Gvozdovskyy I. Electro- and photoswitching of undulation structures in planar cholesteric layers aligned by a polyimide film possessing various values of anchoring energy. Liq Cryst. 2017; xx: 1-17. (Published online: doi:10.1080/02678292.2017.1359691).

[30]. Senyuk BI, Smalyukh II, Lavrentovich OD. Undulations of lamellar liquid crystals in cells with finite surface anchoring near and well above the threshold. Phys Rev E. 2006; 74: 011712-1-13. Doi: 10.1103/PhysRevE.74.011712

[31]. Smalyukh II, Lavrentovich OD. Anchoring-mediated interaction of edge dislocations with bounding surfaces in confined cholesteric liquid crystals. Phys Rev Lett. 2003; 90: 085503-1-4. Doi: 10.1103/PhysRevLett.90.085503

[32]. Rynes EP, Brown CV, Strömer JF. Method for the measurement of the $K_{22}$ nematic elastic constant. App Phys Lett. 2003; 82: 13–15. Doi: 10.1063/1.1534942

[33]. Chigrinov VG, Kozenkov VM, Kwok HS, Photoalignment of Liquid Crystalline Materials: Physics and Applications. England. A John Wiley & Sons, Ltd., Publication, 2008.

[34]. Kozenkov YuV, Chigrinov VG, Kwon SB. Photoanisotropic effects in poly(vinyl-cinnamate) derivatives and their applications. Mol Cryst Liq Cryst. 2004; 409: 251-267. Doi: 10.1080/1542140049043190

[35]. Bryan brown GP, Sage IC. Photoinduced ordering and alignment properties of polyvinylcinnamates. Liq Cryst 1996; 20: 825-829. Doi: 10.1080/02678299608033178

[36]. Voloshchenko D, Khizhnyak A, Reznikov Yu, et al. Control of an easy-axis on nematic-polymer interface by light action to nematic bulk. Jpn J Appl Phys. 1995; 34: 566-571. Doi: 10.1143/JJAP.34.566

[37]. Kumar S, Kim JH, Shi Y. What aligns liquid crystals on solid substrates? The role of surface roughness anisotropy. Phys Rev Lett. 2005; 94: 077803-4. Doi: 0.1103/PhysRevLett.94.077803

[38]. Cull B, Shi Y, Kumar S, et al. Anisotropic surface morphology of poly(vinyl 4-methoxy-cinnamate) and 12-8(poly)diacetylene thin films induced by linear photopolymerization. Phys Rev E: Stat Phys, Plasmas, Fluids, Relat Interdiscip Top. 1996; 53: 3777-3781. Doi: 10.1103/PhysRevE.53.3777

Figure 1. (a) SEM image of the pure MSTL after processing with the NLL method. (b) Dependence of the average depth A on the LPF $\mathscr{I}$ of the pure MSTL after the NLL processing. The cross section of microgrooves of the Ti layer: (c) pure, possessing the average period Λ = 0.92 μm and average depth A = 225 nm and (d) coated with the ODAPI film, having the average period Λ = 0.9 μm and average depth A = 175 nm. In the NLL method the LPF $\mathscr{I}$ and scanning speed $v$ was 0.582 J/cm$^2$ and 1500 mm/s, respectively.

Figure 2. Dependence of the twist angle $\varphi$ of the LC cell on: (a) the scanning speed $v$ at a constant LPF $\mathscr{I}$ = 0.55 J/cm$^2$ and (b) the LPF $\mathscr{I}$ at a constant scanning speed $v$ = 1500 mm/s during the processing of the Ti layer with the NLL method. The LC cell consists of the tested substrate having the pure MSTL (closed "blue" squares and diamonds) and MSTL coated with the ODAPI film (closed "red" circles and triangles). The dashed line is a guide to the eye.

Figure 3. Dependence of the calculated azimuthal AE (Equation 2) of aligning films on the scanning speed $v$ during the processing of the Ti layer with the NLL method. The LC cell consists of the tested substrate having the pure MSTL (closed "blue" triangles) and MSTL coated with the ODAPI film (closed "red" spheres). The dashed line is a guide to the eye.

Figure 4. Dependence of the calculated azimuthal AE (Equation 2) of aligning films on the LPF $\mathscr{I}$ at constant scanning speed $v$: (a) 500 mm/s, (b) 1500 mm/s and (c) 2600 mm/s during the processing of the Ti layer with the NLL method. The LC cell consists of the tested substrate having the pure MSTL (closed "blue" triangles) and MSTL coated with the ODAPI film (closed "red" spheres). The dashed line is a guide to the eye.